\documentclass[12pt]{article}
\usepackage{epsfig}
\usepackage{amssymb}
\def\Title#1{\begin{center} {\Large {\bf #1} } \end{center}}

\begin{document}

\Title{Adler Function in the Analytic Approach to QCD}

\bigskip\bigskip

\begin{raggedright}

{\it A.V.~Nesterenko \\
Bogoliubov Laboratory of Theoretical Physics \\
Joint Institute for Nuclear Research \\
Dubna, 141980, Russian Federation}
\bigskip\bigskip
\end{raggedright}

\centerline{\parbox[t]{5in}{\textbf{Abstract}.
The low energy behavior of the Adler function $D(Q^2)$ is studied by
employing recently derived integral representation for the latter.
This representation embodies the nonperturbative constraints on
$D(Q^2)$, in particular, it retains the effects due to the
nonvanishing pion mass. The Adler function is calculated within the
developed approach by making use of its perturbative approximation as
the only additional input. The obtained result agrees with the
experimental prediction for the Adler function in the entire energy
range and possesses remarkable stability with respect to the higher
loop corrections.}}

\section{Introduction}

The Adler function~$D(Q^2)$~\cite{Adler} plays a crucial role in
various issues of elementary particle physics. Specifically,
theoretical description of such strong interaction processes as
$e^{+}e^{-}$~annihilation into hadrons and inclusive $\tau$~lepton
decay is inherently based on this function. Moreover, Adler function
is essential for confronting precise experimental measurements of
some electroweak observables with their theoretical predictions. In
turn, the latter represents a decisive test of the validity of the
Standard Model and imposes strict restrictions on possible ``new
physics'' beyond it.

The perturbation theory still remains the only reliable tool for
calculating the Adler function at high energies. Namely, in the
asymptotic ultraviolet region~$D(Q^2)$ can be approximated by the
power series in the strong running coupling~$\alpha_{s}(Q^2)$.
However, spurious singularities of the latter, being the artifacts of
perturbative calculations, invalidate this expansion at low energies.
In turn, this significantly complicates theoretical description of
the low--energy experimental data, and eventually forces one to
resort to various nonperturbative approaches.

An important source of the nonperturbative information about the
hadron dynamics at low energies is provided by the relevant
dispersion relations. The latter, being based on the general
principles of the local Quantum Field Theory, supply one with the
definite analytic properties in a kinematic variable of a physical
quantity at hand. The idea of employing this information together
with perturbation theory and renormalization group~(RG) method forms
the underlying concept of the so--called ``analytic approach'' to
Quantum Chromodynamics (QCD)~\cite{ShSol}. Some of the main
advantages of this approach are the absence of unphysical
singularities and a fairly good higher--loop and scheme stability of
outcoming results. The analytic approach has been successfully
employed in studies of the strong running coupling~\cite{ShSol,PRD},
perturbative series for QCD observables (see paper~\cite{APT} and
references therein), meson spectrum~\cite{Prosperi}, chiral symmetry
breaking~\cite{CSB}, and electromagnetic pion form
factor~\cite{Stefanis}.

The primary objective of this paper is to study the infrared behavior
of the Adler function by employing recently obtained integral
representation~\cite{AdlerIR}. The latter has been derived in a
general framework of the analytic approach to QCD, the effects due to
the pion mass being retained. It is also of a particular interest to
examine the stability of the calculated Adler function with respect
to the higher loop corrections.

The layout of the paper is as follows. In Section~2 the dispersion
relation for the Adler function and its interrelation with the
$R$--ratio of electron--positron annihilation into hadrons are
overviewed. In Section~3 a novel integral representation for~$D(Q^2)$
is discussed and the calculation of the Adler function within the
developed approach is presented. In Section~4 the obtained results
are summarized.

\section{The Adler function}

The Adler function $D(Q^2)$~\cite{Adler} naturally appears in the
theoretical description of the process of electron--positron
annihilation into hadrons. Specifically, the measurable ratio of two
cross--sections is proportional to the discontinuity of the hadronic
vacuum polarization function~$\Pi(q^2)$ across the physical cut:
\begin{equation}
\label{RDef}
R(s) = \frac{\sigma\left(e^{+}e^{-} \to \mbox{hadrons}; s\right)}
{\sigma\left(e^{+}e^{-} \to \mu^{+}\mu^{-}; s\right)} =
\frac{1}{2 \pi i} \lim_{\varepsilon \to 0_{+}}
\left[\Pi(s - i \varepsilon) - \Pi(s + i \varepsilon)\right]
\end{equation}
with $s=q^2>0$ being the center--of--mass energy squared. It is worth
noting here that~$R(s)$ vanishes identically for the energies below
the two--pion threshold due to the kinematic restrictions, see also
Ref.~\cite{Feynman}. The mathematical implementation of the latter
condition consists in the fact that~$\Pi(q^2)$ has the only cut
$q^2 \ge 4 m_{\pi}^2$ along the positive semiaxis of real~$q^2$ and
satisfies the once--subtracted dispersion relation~\cite{Adler,Feynman}
\begin{equation}
\label{PDisp}
\Pi(q^2) = \Pi(q_0^2) - \left(q^{2}-q_0^2\right)
\!\int_{4 m_{\pi}^2}^{\infty}\frac{R(s)}{(s-q^2)(s-q_0^2)}\,d s,
\end{equation}
where $m_{\pi} \simeq 135\,$MeV is the mass of the $\pi^{0}$ meson.

For practical purposes it proves to be convenient to deal with the
so--called Adler function~\cite{Adler}, which is defined as
\begin{equation}
\label{AdlerDef}
D(Q^2) = \frac{d\, \Pi(-Q^2)}{d \ln Q^2}
\end{equation}
and, therefore, does not depend on the choice of subtraction
point~$q_0^2$ in the dispersion relation~(\ref{PDisp}). In
Eq.~(\ref{AdlerDef}) $Q^2 = -q^2 \ge 0$ denotes a spacelike momentum.
In addition to the relevance to the strong interaction processes
mentioned in the Introduction, the Adler function~(\ref{AdlerDef})
plays a crucial role for the congruous analysis of hadron dynamics in
spacelike and timelike domains. In particular, the required link
between the experimentally measurable $R$--ratio~(\ref{RDef}) and
theoretically computable Adler function~(\ref{AdlerDef}) is
represented by the dispersion relation~\cite{Adler}
\begin{equation}
\label{AdlerDisp}
D(Q^2) = Q^2 \int_{4 m_{\pi}^2}^{\infty}
\frac{R(s)}{(s + Q^2)^2} \, d s.
\end{equation}
At the same time, one is also able to continue an explicit theoretical
expression for the Adler function~(\ref{AdlerDef}) into timelike domain
by making use of the inverse relation
\begin{equation}
\label{AdlerInv}
R(s) = \frac{1}{2 \pi i} \lim_{\varepsilon \to 0_{+}}
\int_{s + i \varepsilon}^{s - i \varepsilon}\!
D(-\zeta) \, \frac{d \zeta}{\zeta},
\end{equation}
where the integration contour lies in the region of analyticity of
the integrand~\cite{RKP82}.

Although there are no direct measurements of the Adler
function~(\ref{AdlerDef}), it can be restored by employing the data
on $R$--ratio~(\ref{RDef}). Specifically, in the integrand of the
dispersion relation~(\ref{AdlerDisp}) one usually approximates~$R(s)$
by its experimental measurements at low and intermediate energies,
and by its theoretical prediction at high energies. For the energies
below the mass of the $\tau$~lepton the $R$--ratio (which possesses
rather large systematic uncertainties in the infrared domain) can be
substituted (up to the isospin breaking effects) by precise spectral
function of the vector current~$R^{\mbox{\scriptsize
V}}_{\mbox{\scriptsize exp}}(s)$~\cite{ALEPH} extracted from the
hadronic $\tau$~decays. Thus, the $R$--ratio in Eq.~(\ref{AdlerDisp})
can be parameterized by $R(s) = R^{\mbox{\scriptsize
V}}_{\mbox{\scriptsize exp}}(s)\,\theta(s_0-s) +
R^{(3)}_{\mbox{\scriptsize theor}}(s)\,\theta(s-s_0)$, where
$\theta(x)$ is the Heaviside step--function,
$R^{(3)}_{\mbox{\scriptsize theor}}(s)$ stands for the theoretical
prediction of $R(s)$ at three--loop level, $n_f=3$ is assumed, and
$s_0 = 2.1\,$GeV$^2$. The overall factor $N_{c}\sum_{f}Q_{f}^{2}$ is
omitted throughout, where $N_{c}=3$ is the number of colors and
$Q_{f}$ denotes the charge of the quark of the $f$th flavor. Computed
in this way experimental prediction for the Adler function is
presented in Fig.~1 by shaded band, see Ref.~\cite{InPrep} for the
details.

As it has been mentioned above, the high--energy behavior of the
Adler function~(\ref{AdlerDef}) can be approximated by the power
series in the strong running coupling~$\alpha_{s}(Q^2)$ in the
framework of the perturbative approach. Specifically, at the
$\ell$--loop level
\begin{equation}
\label{AdlerPert}
D^{(\ell)}_{\mbox{\scriptsize pert}}(Q^2) = 1 +
\sum\nolimits_{j=1}^{\ell} d_j\left[\alpha^{(\ell)}_{s}(Q^2)\right]^j,
\qquad Q^2\to\infty,
\end{equation}
where $\alpha^{(\ell)}_{s}(Q^2)$ is the $\ell$--loop perturbative
invariant charge. The expansion coefficients~$d_{j}$ are known up to
the three--loop level, in particular, $d_1=1/\pi$. The numerical
estimation~\cite{KatStar} of the uncalculated yet four--loop
coefficient $d_{4}$ is adopted in what follows. However, as one may
infer from Fig.~1, the perturbative approximation~(\ref{AdlerPert})
is reliable for the energies $Q \gtrsim 1.5\,$GeV only. Besides,
expansion~(\ref{AdlerPert}) is incompatible with the dispersion
relation~(\ref{AdlerDisp}) due to unphysical singularities of the
running coupling~$\alpha_{s}(Q^2)$ in the infrared domain. The latter
also causes certain difficulties for processing the low--energy
experimental data.

\section{Novel integral representation for~$D(Q^2)$}

The aforementioned integral relations~(\ref{AdlerDisp})
and~(\ref{AdlerInv}) express the Adler function~(\ref{AdlerDef}) and
$R$--ratio~(\ref{RDef}) in terms of each other. For practical
purposes it proves to be convenient to express both these quantities
in terms of the common spectral function. This objective can be
achieved by employing Eqs.~(\ref{AdlerDisp}) and~(\ref{AdlerInv}),
the parton model prediction $R_{0}(s) =
\theta(s-4m_{\pi}^{2})$~\cite{Feynman}, and the fact that the strong
correction to the Adler function vanishes in the asymptotic
ultraviolet limit $Q^2 \to \infty$. Eventually one arrives at (see
Refs.~\cite{AdlerIR,InPrep} for the details)
\begin{eqnarray}
\label{AdlerInt}
D(Q^2) &=& \frac{Q^2}{Q^2+4m_{\pi}^2} \left[1 + \int_{4m_{\pi}^2}^{\infty}
\rho_{\mbox{\tiny D}}(\sigma)\, \frac{\sigma - 4m_{\pi}^2}{\sigma+Q^2}\,
\frac{d \sigma}{\sigma}\right], \\
\label{RInt}
R(s) &=& \theta(s-4m_{\pi}^2) \left[1 + \int_{s}^{\infty}
\rho_{\mbox{\tiny D}}(\sigma) \frac{d \sigma}{\sigma}\right].
\end{eqnarray}
Here the spectral function $\rho_{\mbox{\tiny D}}(\sigma)$ can be
determined either as the discontinuity of the theoretical expression
for the Adler function across the physical cut $\rho_{\mbox{\tiny
D}}(\sigma) = \mbox{Im}\, D(-\sigma + i0_{+})/\pi$ or as the
numerical derivative of the experimental data on $R$--ratio
$\rho_{\mbox{\tiny D}}(\sigma) = - d\,R(\sigma)/d\,\ln\sigma$. It is
worth noting that Eq.~(\ref{AdlerInt}) embodies the nonperturbative
constraints on Adler function arising from the dispersion
relation~(\ref{AdlerDisp}). Besides, Eq.~(\ref{RInt}) by construction
properly accounts for the effects due to the analytic continuation of
spacelike theoretical results into timelike domain.

In order to compute the Adler function in the framework of the
approach at hand, one first has to determine the spectral
function~$\rho_{\mbox{\tiny D}}(\sigma)$. In what follows we restrict
ourselves to the study of only perturbative contributions to the
latter, namely
\begin{equation}
\label{RhoPert}
\rho^{(\ell)}_{\mbox{\scriptsize pert}}(\sigma) = \mbox{Im}\,
D^{(\ell)}_{\mbox{\scriptsize pert}}(-\sigma + i 0_{+})/\pi,
\end{equation}
where $D^{(\ell)}_{\mbox{\scriptsize pert}}(Q^2)$ is given by
Eq.~(\ref{AdlerPert}). It is worthwhile to mention also that in the
limit of massless pion ($m_{\pi}=0$) the obtained expressions for the
Adler function~(\ref{AdlerInt}) and $R$--ratio~(\ref{RInt}) become
identical to those of the so--called Analytic perturbation theory
(APT)~\cite{APT}, the definition~(\ref{RhoPert}) being assumed.

\begin{figure}[t]
\centerline{\epsfig{file=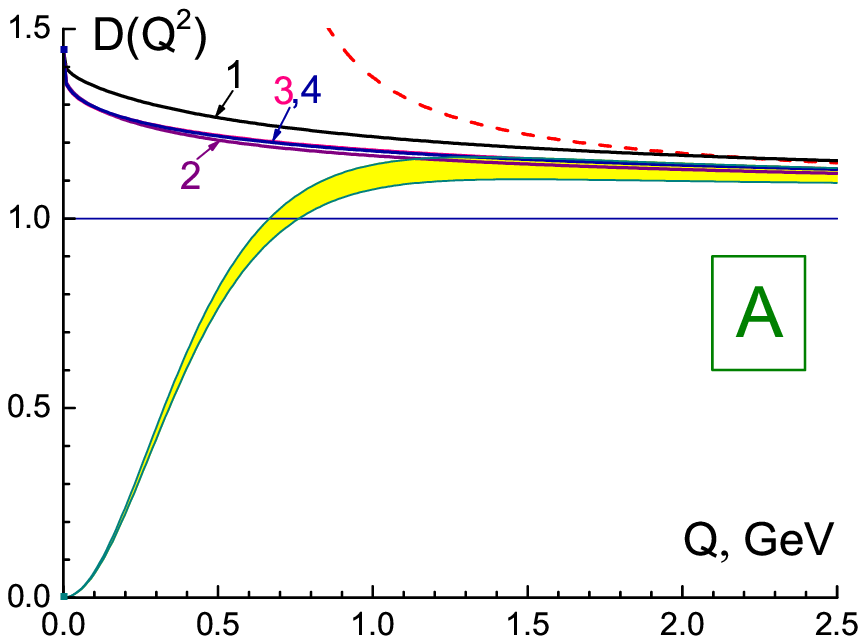, width=70mm}
\hspace{7.5mm}
\epsfig{file=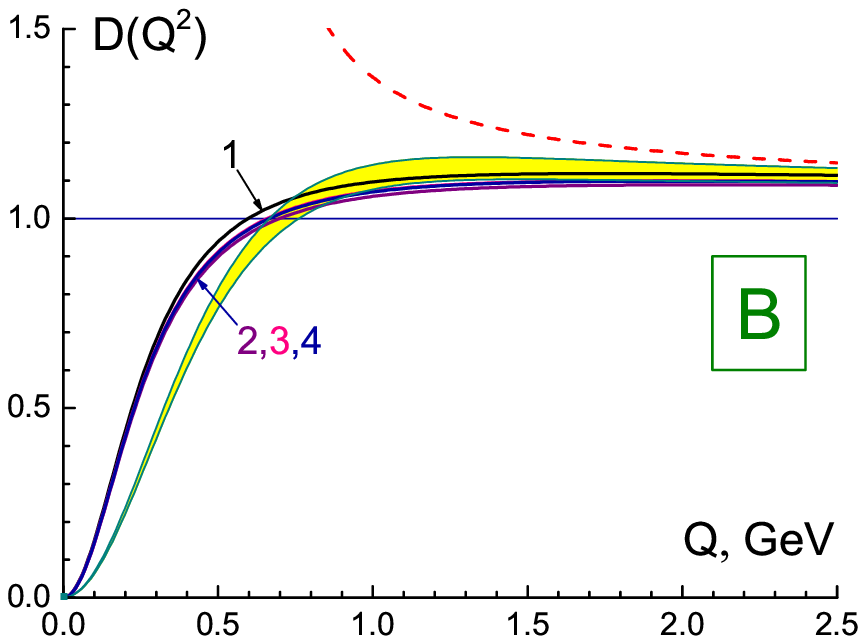,width=70mm}}
\caption{{\small The Adler function~(\ref{AdlerInt}) (solid curves)
calculated by making use of the spectral function~(\ref{RhoPert}) in
the massless~(A) and massive~(B) cases. Numerical labels correspond to
the loop level considered. The experimental prediction for $D(Q^2)$ is
shown by the shaded band, whereas its perturbative approximation is
denoted by the dashed curve.}}
\label{fig:Adler}
\end{figure}

For the illustration of the significance of the pion mass within the
approach at hand, it is worth presenting the Adler
function~(\ref{AdlerInt}) computed by making use of the spectral
function~(\ref{RhoPert}) for both, massless and massive cases. The
obtained results are presented in Fig.~1 by solid curves. In the case
of the massless pion (which is identical to the APT~\cite{APT}), one
arrives at the result, which is free of infrared unphysical
singularities, but fails to describe the Adler function for the
energies $Q \lesssim 1.0\,$GeV, see Fig.~1$\,$A. It is worth noting
here that in the framework of the massless APT the infrared behavior
of $D(Q^2)$ can be further improved by additionally invoking into
consideration relativistic quark mass threshold
resummation~\cite{MSS} or vector meson dominance
assumption~\cite{Cvetic}.

At the same time, as one may infer from Fig.~1$\,$B, for the case of
the nonvanishing pion mass the representation~(\ref{AdlerInt}) is
capable of providing an output for the Adler function, which agrees
with its experimental prediction in the entire energy
range~\cite{AdlerIR}. Moreover, the Adler function~(\ref{AdlerInt})
is remarkably stable with respect to the higher loop corrections.
Namely, the relative difference between the $\ell$--loop and
$(\ell+1)$--loop expressions for $D(Q^2)$~(\ref{AdlerInt}) is less
than $4.9\%$, $1.5\%$, and $0.3\%$ for $\ell=1$, $\ell=2$, and
$\ell=3$, respectively, for $0 \le Q^2 < \infty$, see
Ref.~\cite{InPrep} for the details. It is worthwhile to mention also
that the obtained results are supported by recent studies of meson
spectrum in the framework of the Bethe--Salpeter
formalism~\cite{Prosperi}.

\section{Summary}

The infrared behavior of the Adler function is examined by employing
representation~(\ref{AdlerInt}), which accounts for the pion mass
effects. The approach at hand possesses all the appealing features of
the massless APT~\cite{APT}. Namely, it supplies a self--consistent
analysis of spacelike and timelike experimental data; additional
parameters are not introduced into the theory; the outcoming results
possess no unphysical singularities and display enhanced higher loop
stability. In addition, the developed approach provides a reliable
description of the Adler function in the entire energy range.

\medskip
Partial support of grants RFBR 05-01-00992 and NS-5362.2006.2 is
acknowledged.

\end{document}